\documentclass[mathleft,final]{an}
\usepackage{graphicx}
\usepackage{times}
\begin{document}

\Pagespan{1}{}
\Yearpublication{2010}%
\Yearsubmission{2009}%
\Month{1}%
\Volume{}%
\Issue{}%

\title{SSS in young stellar populations and the ``prompt" component of Type Ia Supernovae}

\author{Thomas Nelson\inst{1,2}\fnmsep\thanks{Corresponding author:
  \email{thomas.nelson@nasa.gov}\newline}
\and  Marina Orio\inst{3,4}
\and  Francesco Di Mille\inst{5}
}
\titlerunning{Young SSS in M31}
\authorrunning{T. Nelson, M. Orio \& F. De Mille}
\institute{
NASA Goddard Space Flight Center, Greenbelt, MD 20771, USA
\and 
University of Maryland, Baltimore County, 1000 Hilltop Circle, Baltimore MD 21250, USA
\and 
Dept. of Astronomy, University of Wisconsin-Madison, 475 N. Charter St, Madison, WI 53706, USA
\and
INAF-Osservatorio Astronomico di Padova, vicolo Osservatorio 5, I-35122 Padova, Italy
\and
Dipartimento di Astronomia, Universita di Padova, vicolo Osservatorio, 2, I-35122, Padova, Italy}
\received{30 May 2005}
\accepted{11 Nov 2005}
\publonline{later}


\abstract{We present the results of a search for UV and optical counterparts of the SSS
population in M31.  We find that out of the 56 sources we included in our search, 16 are
associated with regions of ongoing or recent star formation.  We discuss two particularly interesting sources that are identified optically as early type stars, one of which displayed long term X-ray evolution similar to that observed in classical novae.  We discuss the physical origin of supersoft X-rays in these and the other SSS in young regions, and their possible link to the so-called ``prompt" component of the Type Ia supernova population.}

\maketitle

\section{Introduction}
The progenitors of type Ia supernovae (SNe Ia) remain elusive, despite the ever growing number of observed events.  Currently favoured progenitor models consist of white dwarfs (WDs) in binary systems which grow in mass until they reach the Chandrasekhar limit and explode.  However, the details vary from model to model.  It is clear that for any model to succeed, it must be able to account for the observed properties of the SNe Ia population, including the lack of detectable hydrogen in their spectra, the variations in maximum brightness, and the appearance of SNe Ia in both young (the so-called ``prompt" component) and old (``tardy") stellar populations (e.g. Scannapieco \& Bildsten 2005).  

Ever since van den Heuvel et al. (1992) proposed the nuclear shell-burning accreting WD model to explain the observed characteristics of supersoft X-ray sources (SSS), they have attracted a great deal of interest as possible progenitors of SNe Ia (see e.g. Di Stefano 1996; Starrfield et al. 2004; Nomoto et al. 2008).  In principle, steady hydrogen burning allows the WD to grow in mass over time as the helium rich ashes build up on its surface.  If both the initial mass and the mass accretion (growth) rate are high enough, the WD can reach the Chandrasekhar limit and explode as a SN Ia.  Based on estimates of the numbers of these steady burning WDs, population synthesis models can predict the number of type Ia supernovae that will result, and compare this to the observed rate.

M31 hosts the largest population of SSS of any galaxy (over 80 sources), and in recent years has been the target of a number of surveys at different wavelengths.  Despite the relative proximity of M31, the properties of SSS make direct detections of optical counterparts very difficult.  Most of the Magellanic Cloud SSS counterparts have very blue colours (typical values are U-B $\sim$ -1 and B-V $\sim$ -0.4, see the review by \v{S}imon 2003 and references therein.)  The V band magnitudes of LMC sources range from $\sim$21.7 (RX J0439.8-6809, van Teeseling, Reinsch and Beuermann 1993) to $\sim$16.7 (RX J0513.9-6951 in high optical state, Alcock et al. 1996).  Thus, at the distance of M31 (780 kpc, Stanek and Garnavich 1998) these sources would have observed magnitudes between 28 and 23 mags.  This makes them difficult to observe from all but the largest ground based observatories.

Popham \& Di Stefano (1996) showed that if SSS are WDs experiencing shell nuclear burning, then emission from the burning region cannot by itself account for the observed optical flux.  Instead, the primary source of optical flux in these systems must be the accretion disk which is irradiated by, and reprocesses, the energy released by the burning layer.  Their models of such irradiated disks predict that the flux in the ultraviolet is 10--50 times larger than that in the optical (2.5--4 magnitudes).  This has been confirmed observationally by UV spectroscopy of several Magellanic Cloud sources (e.g. CAL 83 and RX J0513.9-6951, G\"{a}nsicke et al. 1998).  The flux from the shell burning, if present, also peaks in the EUV (e.g. Fujimoto 1982).  Therefore, it makes sense to conduct a search for the counterparts of the M31 SSS population in the UV, where they are brighter and more likely to be detected.  In this paper we describe the results of such a search.

\section{Data}
We used the Galactic Evolution Explorer (\textit{GALEX}) images of M31 to search for UV counterparts of SSS.  The data were originally obtained in 2003 as part of the Nearby Galaxies Survey (NGS, Thilker et al. 2005), and in 2007 during the Deep Imaging Survey (DIS).  These data are ideal for our purposes since they provide imaging of approximately constant depth over the whole galaxy at two epochs, and in two energy bands (FUV, 1350--1800 $\mathring{A}$ and NUV, 1800--2800 $\mathring{A}$). \textit{GALEX} has a moderate spatial resolution of 3.3" (5") in the FUV (NUV) bands.

To compliment the UV images, we also used the data of the Local Group Survey of M31 (Massey et al. 2006).  This dataset is comprised of U, B, V, R and I band images of the entire disk of M31 down to $\sim$23 magnitude, and a photometric catalogueue with 1\% photometry at 21 mags, and $<$10\% photometry at 23 mags.  The optical data are useful for determining the nature of the UV sources seen in the \textit{GALEX} data, and for characterizing the environment of SSS in general (see Orio et al. 2009, submitted).

\section{Search results}
We created an updated catalogueue of SSS in M31 including sources previously published in the literature (Supper et al. 1997; Kahabka 1999; Greiner et al. 2004; DiStefano et al. 2004; Pietsch et al. 2005; Orio 2006; Stiele et al. 2008) and 6 new sources identified by us for the first time in more recent X-ray images (Orio et al. 2009).  Excluding SSS already identified as classical novae, our catalogueue contains 64 individual sources.  Only three sources identified as SSS previously are excluded; source 518 of Pietsch et al. (2005), which we exclude due to its low S/N detection, r3-122 (Di Stefano et al. 2004) which we identify as a foreground V=13.36 star, and r1-25 (Di Stefano et al., 2004) which is observed primarily in a hard state in more recent observations.  Due to the limited spatial resolution of \textit{GALEX}, and problems with source confusion and diffuse light in the core of M31, we excluded 8 SSS lying within 5 arcmin of the galaxy centre. This left us a total of 56 SSS for our counterpart search. For each source, we created a circular region with radius equal to the 2$\sigma$ positional error from the original reference and added an additional 1" in quadrature to account for the 2$\sigma$ systematic offset in the \textit{GALEX} WCS.  We then searched all UV images for sources falling within these regions,  and consider any detections potential counterparts of the X-ray sources.

Of the 56 included SSS, only 16 have candidate UV counterparts.  The majority of these 16 SSS were discovered using \textit{ROSAT}, and therefore have much larger positional uncertainties than sources with either \textit{XMM-Newton} or \textit{Chandra} detections.  In many cases multiple, unresolved UV sources lie within the positional error circle, and in fact only 3 objects are sufficiently resolved to do photometry.  For those sources, we used the apphot IRAF routine (part of the the digiphot package) to perform the photometry.  Source counts were extracted from a circular aperture with a 6 pixel radius, and background counts from a larger annular aperture of inner radius 12 pixels, and width 5 pixels. We then converted the resulting count rates to AB magnitudes using the conversion factors outlined in the \textit{GALEX} observers handbook.  To facilitate comparison with theoretical isochrones, we converted these AB magnitudes to the Vega system by subtracting 2.224 (1.699) mags from the measured FUV (NUV) AB system value.

The results are plotted in Figure 1.  We show the observed magnitudes (i.e. no reddening correction), and theoretical isochrones for solar metallicity populations of ages 10$^{7}$--10$^{8}$ yrs (Girardi et al. 2002).  The solid and dashed lines correspond to E(B-V) of 0 and 0.5 respectively (the latter value was chosen as a typical maximum extinction towards the sources in our catalogue, as determined from HI radio maps---see Orio et al. 2009).  Two of the sources have UV colours and magnitudes consistent with young massive stars.  The third source, RX J0040.0+4100 lies to the left of the isochrones, indicating a higher temperature.  This source may have UV characteristics consistent with an accreting, shell burning WD.  The lack of current X-ray detections may indicate that it has transitioned into a UV dominated state, as is observed in CAL 83 (Greiner \& Di Stefano 2002) and RX J0513.9-6951 (Reinsch et al. 2000).

\begin{figure}
\includegraphics[width=3in]{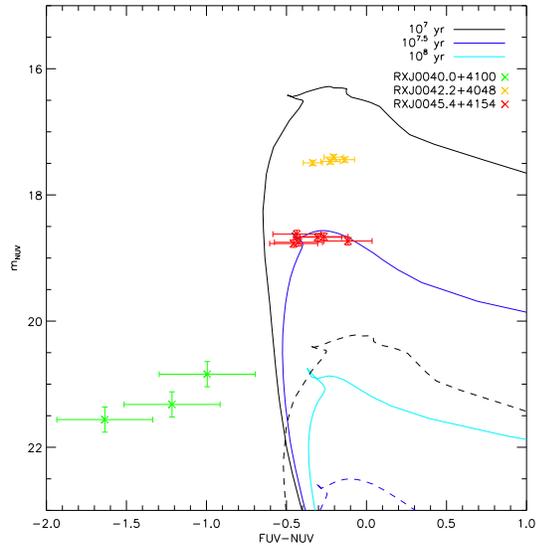}
\caption{\textit{GALEX} magnitudes and colours of three isolated counterparts.  The solid and dashed lines are Padova isochrones for E(B-V) = 0 and 0.5, for ages 10$^{7}$ , 10$^{7.5}$ and 10$^{8}$ yrs.}
\label{f1}
\end{figure}

It is likely that most of the UV sources we find are early type stars since the brightest objects in the \textit{GALEX} filters are stars with temperatures in the range 20000-30000 K.  There are $\sim$30000 UV sources in the GALEX images of M31, mostly concentrated in a annulus around the center (the so-called ring of fire) of area $\sim$1 deg$^{2}$, and thus they have a surface density of $\sim$30000 deg$^{-1}$.  Therefore in a circular region with radius 30" (typical of the 3$\sigma$ positional uncertainty of the \textit{ROSAT} sources) we expect $\sim$7 UV sources.  This means that many of the UV sources are chance alignments.  However, in the following section we discuss two sources which are uniquely identified with early type stars, indicating a physical connection between the SSS and young stellar populations.

\section{Two interesting sources: RX J0042.2+4048 and XMMU J003910.4+404521}
RX J0042.2+4048 is one of only three persistent SSS in M31 (i.e. systems which are always on and always soft). The source is marginally detected off axis in a number of \textit{Chandra} observations, allowing us to refine the X-ray coordinates to within 2".  This position coincides not only with the \textit{GALEX} source discussed above, but also a V = 19.419 star in the LGS data.  The star is very blue, with all LGS colours negative, typical of a young massive star.  

We obtained a spectrum of the optical source with the DOLORES spectrograph at the 3.58 m Telescopio Nazionale Galileo (TNG) of the Italian National Institute of Astrophysics (INAF) using the LR-B grism and a 1 arcsec width slit.  The resulting spectrum is shown in Fig 1.  Overplotted in red is an O9 V star, the closest match from the stellar spectral library of Jacoby et al. 1984. The source matches the template well, except below 4000 \AA, where the signal to noise ratio is low.  The inferred M$_{V}$ is consistent with the observed magnitude.  There is no detection of SII, ruling out a SNR remnant interpretation for this object.  H$\alpha$ is detected in emission at several positions along the slit, suggesting the line arises in a foreground/background HII region.  However, we cannot rule out a contribution from the star itself with the current data, in which case the star would be classified as spectral type Oe.

\begin{figure}
\includegraphics[width=3in]{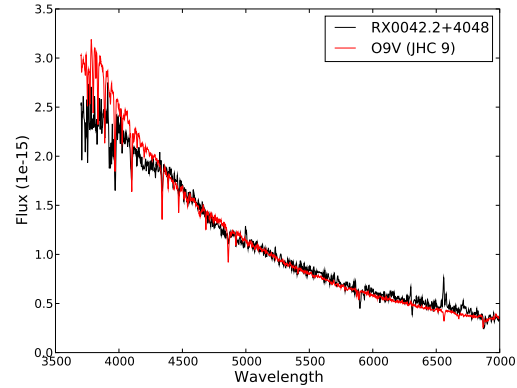}
\caption{TNG spectrum of candidate counterpart of RX J0042.2+4048 (\textit{in black}), with O9 V spectrum overlayed (\textit{in black}).}
\label{f2}
\end{figure}

The other interesting SSS is presented here for the first time.  The source, XMMU J003910.4+404521, was discovered by us in some recently public \textit{XMM-Newton} images.  It is first detected in an image from 2006-07-23, with an EPIC-pn count rate of $\sim$0.08 c/s.  The source was not detected with \textit{ROSAT}, and unfortunately the source position was not previously observed with either \textit{XMM-Newton} or \textit{Chandra} before this date, so constraining a turn on time is not possible.  In subsequent \textit{XMM-Newton} images the source faded, on timescales similar to some classical novae after outburst  We present in Fig 3 the longterm EPIC-pn lightcurve of the source.  Extracting the spectrum from the first observation (see Fig 3 inset plot), we find that it can be fit by an absorbed blackbody model with T$_{\mathrm{eff}}$ = 4--6 $\times$ 10$^{5}$ K and N(H) = 0.7--1.5 $\times$ 10$^{21}$ cm$^{-2}$.  This gives an unabsorbed luminosity in the 0.2--10 keV range of 4 $\times$ 10$^{37}$--10$^{38}$ erg s$^{-1}$, typical of other SSS. We obtained a TOO observation of the source with \textit{Swift} on 2008-05-23 which resulted in a non-detection, with an upper limit of L$_{X}$ $<$ $\sim$10$^{36}$ erg s$^{-1}$. 

\begin{figure}
\includegraphics[width=3in]{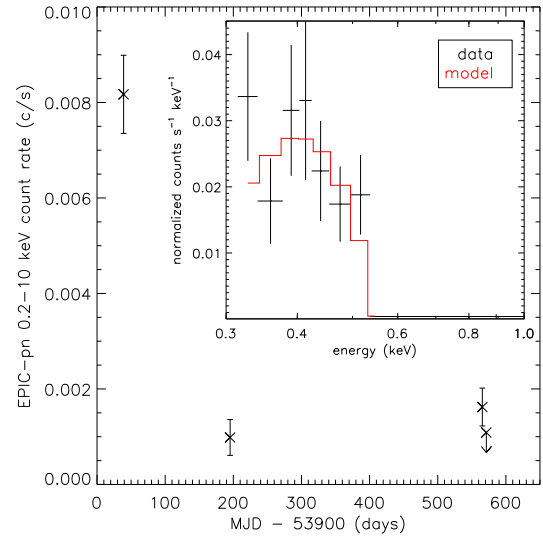}
\caption{Long term EPIC-pn lightcurve of XMMU J003910.4+404521.  The spectrum of the first observation, along with the best fit blackbody model, is shown in the inset plot.}
\label{f3}
\end{figure}

Most interestingly, XMMU J003910.4+404521 is optically identified with a V = 21.827 mag star in the LGS catalogue of M31.  This is much too bright to be a classical nova, which at the distance of M31 have quiescent V band magnitudes of $\sim$29 mags.  Although some recurrent novae have giant companions and can therefore be brighter, the observed LGS colours of this system are very blue, with B-V = -0.33 and U-B = -0.71.  In fact, these colours are much more typical of a young, B type star.  We can rule out that the optical data were obtained shortly after a missed nova outburst since the source is detected at the same brightness level 4 years later in our \textit{Swift} UV Optical Telescope images.  The positive value of V-R (0.15) and the negative value of R-I (-0.17) indicate an enhancement in the R band which could be explained by the presence of H$\alpha$ in emission, making this a candidate Be star.  We also note that the B-V colour is bluer than that predicted by stellar evolution models (even without a reddening correction), but is similar to the values observed in SSS in the Magellanic Clouds where the source of the light is the accretion disk around the white dwarf.  

The probability that these optical identifications are chance alignments is small.  Using the reddening free Q parameter\footnote{$Q = (U-B)-0.72(B-V)$}, we selected $\sim$80000 stars with spectral type earlier than B6 (Q $<$ -0.67) from the LGS catalogue with M$_{V}$ $<$ 23 mags.  Again, most of these stars are located in the ``ring of fire" annular region and so the surface density of these stars is $\sim$80000 deg$^{-1}$. We estimate the probability of a chance alignment within a 2" radius circular region to be $<$8\%.

\section{The nature of young SSS and their connection to the ``prompt" SNe Ia}
Two other SSS in the LMC and SMC have also been identified with Be stars (Kahabka et al. 2006, Takei et al 2008).  Although early type stars are known X-ray emitters, even the brightest sources (early O stars with Lx $\sim$10$^{33}$ erg s$^{-1}$, Cohen 2000) are too faint to be detected in M31.  High mass X-ray binaries (early type stars with neutron star or black hole companions) could be detected in outburst (L$_{X}$ up to 10$^{38}$ erg s$^{-1}$), but have harder X-ray emission (e.g. Eger \& Haberl 2008).  The luminosities and spectral characteristics of the two sources highlighted here are typical of shell burning WDs.  The long-term X-ray behaviour of XMMU J003910.4 +404521 is even reminiscent of classical novae in outburst.  Could these young systems harbour shell burning WDs?   This may not be as unlikely as it first appears.  Stars with initial masses $\leq$10 M$_{\odot}$ are understood to end their evolution as WDs, and a number of late O and early B type stars with WD companions have now been found (e,g, y Pup, $\theta$ Hya and 16 Dra).  This implies that the WD progenitor was of earlier spectral type. However, none of the systems discovered to date have small enough orbital separations to allow mass transfer onto the WD.

The possible Oe/Be nature of these counterparts opens up another evolutionary path leading to accreting WDs in young systems.  These systems are B (or more rarely late O) type stars which show emission lines in their spectra, thought to arise in a circumstellar disk of material lost from their equators. They also rotate extremely rapidly, often at speeds close to the critical breakup velocity (see the review in Porter \& Rivinius 2003).  Why Oe/Be stars rotate so rapidly has been the source of much debate, but the existence of Be-NS binaries reveals that some fraction of systems are likely spun up as the result of mass transfer during close binary evolution.  Studies of open clusters in the Southern Hemisphere by McSwain et al (2005) show that up to 73\% of Be stars could have been spun up by binary evolution. 

Rappaport and van den Heuvel (1982) showed that Be-NS binaries can evolve from binaries with initial primary mass greater than 10 M$_{\odot}$.  However, in systems with a less massive primary, white dwarfs or helium stars are predicted instead.  Raguzova (2001) carried out evolutionary studies of Be stars formed by close binary evolution, and determined that 70\%  will have WD companions, of which 58\% will be of CO composition.  The masses of these WDs peak at a higher value (0.8 M$_{\odot}$) than in the field because of the larger progenitor masses.  Raguzova also found that the binary orbits will be circular for Be-WD systems (no momentum kick since no supernova) and that the WD should generally lie \textit{inside} the Be star disk.  Therefore, it is conceivable that these WDs will accrete matter from the Be star disk and experience episodes of shell burning (as in nova systems), which would explain the transient supersoft X-rays.

If these systems can grow in mass, they may reach the Chandrasekhar limit and explode as SNe Ia.  Since these binaries will form within 5 $\times$ 10$^{7}$ yrs after a starburst, and their WDs are initially massive, they may accrete enough material to explode in a short time ($<$ 10$^{7}$ yrs).  Therefore, these systems could account for the ``prompt" component of SNe Ia which occur within 10$^{8}$ yrs after an episode of star formation.  Further modelling and observations will be required to fully explore this possibility.

\end{document}